\documentclass[prb,twocolumn,showpacs]{revtex4}
\usepackage{amssymb}
\usepackage{graphicx}
\usepackage{amsfonts}
\usepackage{amsmath}

\begin{document}

\title{Memory effects in adiabatic quantum pumping with parasitic nonlinear dynamics}

\author{F. Romeo and R. Citro}
\affiliation{$^{1}$Dipartimento di Fisica ''E. R. Caianiello''
and Institute CNR-SPIN, Universit{\`a} degli Studi di Salerno, Via S. Allende,
I-84081 Baronissi (Sa), Italy}

\begin{abstract}
The charge current adiabatically pumped through a mesoscopic region coupled to a classical variable obeying a nonlinear dynamics is studied within the scattering matrix approach. Due to the nonlinearity in the dynamics of the variable, an hysteretic behavior of the pumping current can be observed for specific characteristics of the pumping cycle. The steps needed to build a quantum pump working as a memory device are discussed together with a possible experimental implementation.
\end{abstract}



\maketitle
\section{Introduction}
\label{sec:introduction}
The quantum pumping proposed by Thouless\cite{thouless} is a phase coherence effect able to pump dc current by the out-of-phase adiabatic modulation of two system parameters. The phase difference $\varphi$ of the external signals produces a charge current  proportional to $\omega \sin(\varphi)$, $\omega$ being the adiabatic pumping frequency.
After
the pioneering work of Thouless, a scattering matrix approach to the adiabatic quantum pumping has been proposed by Brouwer\cite{brouwer}. Following the Brouwer formulation, the notion of non-interacting quantum pump\cite{non-int-pump, moskalets&buttiker02} has been developed in several directions. The most important application is the possibility to achieve a pure spin current injector, which is an important goal in spintronics.
Recently the idea of interacting quantum pump\cite{int-pump} beyond the free electrons model, has been introduced.
To get read of non-trivial interaction effects, typically described in the context of the Anderson model\cite{anderson61} for quantum-dot systems, a non-equilibrium Green's functions (NEGF) theory of the quantum pumping has been developed using several approximations.
Alternatively to the NEGF approach, a powerful diagrammatic technique based on a generalized master equation method\cite{konig_diagrammatic} has been developed.  More recently, another interesting
study\cite{silva_08} was performed aiming at generalizing
Brouwer's formula for interacting systems to include inelastic
scattering events.\\
 From the experimental\cite{exp-pump} point of view pumping effects have been analyzed in confined nano-structures, as quantum dots, where the realization of a periodic time-dependent potential can be achieved by modulating gate voltages.\\
Very recently it has been proposed the idea that the presence of a classical state variable inside the scattering region of the system can strongly affect the form of the pumped current. This effect is produced by the linear dynamics of a classical variable which modifies the phase relations introducing a dynamical phase shift\cite{deformable-qd}. \\
In this work we study the charge current pumped through a mescoscopic region coupled to a parasitic classical variable governed by a weakly nonlinear dynamics supposed adiabatic on the typical time scales of the electron transport.\\
The organization of the paper is the following: In
Sec.\ref{sec:scattering-theory}, we summarize the scattering matrix
approach to the quantum pumping as formulated
by M. Moskalets and M. B\"{u}ttiker in Ref.[\onlinecite{moskalets&buttiker02}].
In Sec.\ref{sec:pump-non-linear} we derive a general expression for the current pumped through a mesoscopic region containing a parasitic internal degrees of freedom governed by weakly nonlinear dynamics. In Sec.\ref{sec:results} the specific case of the Duffing nonlinear oscillator is discussed, while the conclusions are given in Sec.\ref{sec:conclusions}.

\section{Scattering Theory of Quantum Pumping}
\label{sec:scattering-theory}

In the standard theory of the quantum pumping (i.e. without internal classical dynamics) a mesoscopic scatterer coupled to two external leads (left/right) taken at the same temperature $T$ and electrochemical
potential $\mu \approx E_F$, $E_F$ being the Fermi energy, is considered. When the scatterer is subjected to two controllable parameters $X_i$, whose adiabatic time dependence is  $X_i(t)=X^{0}_i+X^{\omega}_i\sin(\omega t+\varphi_i)$, the mesoscopic phase
coherent sample is characterized by the scattering
matrix $\mathcal{S}(X_1(t),X_2(t))$. Exploiting an instant scattering
description and assuming that the pumping
amplitudes $X^{\omega}_i$ are small quantities compared to
$X^{0}_i$, the scattering matrix
takes the following form:
\begin{equation}
\label{eq:s-expansion}
\mathcal{S}(X_1(t),X_2(t))\approx
\mathcal{S}(X^{0}_1,X^{0}_2)+s_{+}e^{i\omega t}+s_{-}e^{-i\omega
t},
\end{equation}
where $s_{\eta}$ are related to the parametric derivatives of the scattering matrix.
Under equilibrium condition of the external leads, the electrons with energy $E$
entering the scattering region (\textit{in}-states) are described
by the Fermi distribution
$f_\alpha^{in}(E)=f_0(E)=(1+\exp(\frac{E-\mu}{K T}))^{-1}$, where
$\alpha \in \{left, right\}$. On the other hand, the distribution function
$f_{\alpha}^{out}(E)$ of the outgoing particles leaving the
mesoscopic region and entering the reservoir (namely
\textit{out}-states) is affected by the interaction with the
oscillating scatterer allowing the electrons to absorb or emit
an energy quantum $\hbar \omega$. The latter mechanism changes the initial distribution
function $f_0(E)$ and produces a non-equilibrium distribution responsible for the generation of a finite particles current.
To compute the non-equilibrium
distribution of the outgoing particles $f^{out}_{\alpha}(E)$ one introduces two kinds of operators: The fermionic field operator
$\hat{a}_\alpha$ which annihilates an incoming state in the lead
$\alpha$ and the annihilation operator of the outgoing state
$\hat{b}_\alpha$ in the same lead. The operators $\hat{a}_\alpha$ and $\hat{b}_\alpha$ written in energy representation are related to the scattering matrix by the relation
\begin{equation}
\label{eq:scattering}
\hat{b}_\alpha(E)=\sum_\beta
\mathcal{S}^{0}_{\alpha\beta}\hat{a}_\beta(E)+\sum_{\beta,\eta=\pm}s_{\eta,\alpha\beta}\hat{a}_\beta(E+\eta
\hbar \omega ),
\end{equation}
where
$\mathcal{S}^{0}_{\alpha\beta}=\mathcal{S}(X^{0}_1,X^{0}_2)_{\alpha\beta}$.
Exploiting Eq.(\ref{eq:scattering}) the non-equilibrium
distribution $f^{out}_\alpha(E)$ can be computed as
$\langle\hat{b}^{\dagger}_\alpha(E)\hat{b}_\alpha(E)\rangle$,
being $\langle...\rangle$ the quantum-statistical average.
According to this, we obtain
\begin{equation}
f^{out}_\alpha(E)=\sum_\beta
|\mathcal{S}^{0}_{\alpha\beta}|^2f_0(E)+\sum_{\beta,\eta=\pm}|s_{\eta,\alpha\beta}|^2f_0(E+\eta
\hbar \omega ),
\end{equation}
where we have used the relation
$\langle\hat{a}^{\dagger}_{\alpha}(E)\hat{a}_{\beta}(E')\rangle=\delta_{\alpha\beta}\delta(E-E')f_0(E)$.
If we define as positive the current $I_\alpha$ flowing from the
scatterer to the lead $\alpha$ we can write
\begin{equation}
I_\alpha=\frac{e}{h}\int^{\infty}_0
dE[f_{\alpha}^{out}(E)-f_0(E)].
\end{equation}
The above formula can be explicitly evaluated in the small $\omega$ limit and the expression
of the current $I_{\alpha}$ becomes:
\begin{equation}
I_{\alpha}=\frac{e\omega}{2\pi}\sum_\beta(|s_{-,\alpha\beta}|^2-|s_{+,\alpha\beta}|^2).
\end{equation}
In case of a quantum pump working with two time modulated parameters, the dc-current
$I_\alpha$ can be written in the following final form:
\begin{equation}
\label{eq:pump_curr_0}
I_\alpha=\frac{e\omega\sin(\varphi_2-\varphi_1)X^{\omega}_1
X^{\omega}_2}{2\pi}\sum_\beta \Im
\Bigl\{\Bigl(\partial_{X_1}\mathcal{S}^{\ast}_{\alpha\beta}\partial_{X_2}\mathcal{S}_{\alpha\beta}\Bigl)_{0}\Bigl\}.
\end{equation}
When the quantum pumping is performed by $N$ parameters (\textit{multi-parametric case}) with time modulation $X_n(t)=X^{0}_n+X^{\omega}_n\sin(\omega t+\varphi_n)$, the charge current contains terms proportional to $X^{\omega}_nX^{\omega}_m\sin(\varphi_m-\varphi_n)$
showing the sensitiveness of the quantum pumping to all phase differences experienced by the carriers inside the scattering region.

The latter observation is very important in describing the case of an effective multi-parametric case. In this light when a two-parameters Thouless pumping is performed, the dynamics of a parasitic classical state variable describing some degree of freedom of the scatterer can be excited by the external modulations. The effect of such parasitic state variable is the introduction of an internal (\textit{a priori} unknown) dynamics forced by the pumping cycle and acting as a third pumping parameter (effective multi-parametric case). In analogy with the multi-parametric pumping discussed before, we thus expect additional contributions to the pumping current related to the parasitic dynamics. To be precise, in our language a parasitic degree of freedom is a classical state variable related to the internal dynamics (\textit{a priori} unknown) of the scattering region. The notion of parasitic variable is conceptually similar to the one of \textit{parasitic impedance} in electrical circuits theory: i.e. a degree of freedom not considered in designing the device whose dynamics may affect the performance of the system in an unexpected (usually unwanted) way. In that context, once the presence of an \textit{hidden impedance} is recognized, an equivalent circuital model of the whole system can be obtained and a proper control of the device is achieved. In full analogy to this well known situation, we would like to build a scattering theory of the two-parameters Thouless pumping able to take into account the parasitic dynamics. The resulting theory is equally applicable when the internal dynamics is \textit{a priori} known as is the case for a deformable quantum dot system as considered in Ref.[\onlinecite{deformable-qd}].\\
Thus, when a classical state variable is indirectly excited by the pumping parameters, an internal phase shift may activate additional contributions to the charge current. This mechanism becomes very interesting in the case of a weakly nonlinear dynamics able to produce non-trivial internal phase shifts.\\

\section{Quantum Pumping with a parasitic Nonlinear Variable}
\label{sec:pump-non-linear}
We now consider the case when the dynamics of the oscillating scatterer is affected by the internal classical dynamics of a parasitic variable $y(t)$.
The nature of the variable $y$, if not recognized and included in the pumping model, does not allow to precisely control the system even though the external pumping parameters $X_i(t)$ act as a driving force on the parasitic dynamics.
In this way the scattering matrix $\mathcal{S}(X_1(t),X_2(t);y(t))$ depends parametrically on the pumping parameters and on the parasitic variable whose dynamics is controlled by the pumping cycle in a way which is \textit{a priori} unknown. When the equation of motion of $y(t)$ is linear (or when it can be linearized around the working point), the dynamics of the parasitic variable in the presence of external parameters can be described in terms of the classical Green's function $\chi(t-t')$,
\begin{equation}
\label{eq:linear-dynamics}
y(t)=\int_{-\infty}^{\infty} dt' \chi(t-t')F(X_1(t'),X_2(t')),
\end{equation}
$F(X_1,X_2)$ being the forcing term of the differential equation. Once the dynamics has been formally solved the charge current can be computed as done in Eq.(\ref{eq:pump_curr_0}) or in the multi-parametric case. In the latter case, already discussed in Ref.[\onlinecite{deformable-qd}], the dynamical phase shift $\phi_D$ induced by the Fourier transform $|\chi|e^{i\phi_D}$ of $\chi(t-t')$ plays an important role in generating additional contributions to the pumping current. Such additional terms, weighted by $|\chi|$,  can give information on the differential operator related to the parasitic dynamics configuring the quantum pumping as a probe of internal dynamics.\\
When the internal variable follows a nonlinear dynamics all the above considerations are still valid even though new features appear. We consider here the simplest nonlinear dynamics in which the interaction between the electronic degrees of freedom and the parasitic variable is governed by the equation of motion of a forced Duffing oscillator\cite{nota1} (in dimensionless units):
\begin{equation}
\label{eq:duffing_eom}
\ddot{y}+\beta \dot{y}+y+\epsilon[-a y+b y^3+\sum_{i=1,2}g_iX_i^{\omega}\sin(\tau +\varphi_i)]=0,
\end{equation}
where the $g_i$ are gain functions, $a$ is a detuning term able to modify the resonance frequency of the system, while $b>0$ is the nonlinear term\cite{nota2}. Furthermore, the time $\tau$ is normalized to the resonance frequency $\Omega_0$ of the harmonic oscillator (obtained for $\epsilon=0$) while the pumping frequency $\omega$ is tuned on resonance (i.e. $\omega=\Omega_0$). In the weak interaction limit ($\epsilon \ll 1$) one can measure the linear damping in terms of $\epsilon$ making the substitution $\beta\rightarrow \epsilon \beta$. In this way a weakly nonlinear system is obtained which can be treated within the two-timing perturbation theory\cite{strogatz_book94}. Namely two time scales are introduced, one is the fast time scale $t=\tau$ and the other is the slow time scale $T=\epsilon\tau$, while the approximate solution of the problem is written as $y(t, T)\approx y_0(t, T)+\epsilon y_1(t, T)$. The derivative  with respect to $\tau$, i.e. $d/d\tau$, has to be substituted by the operator $\partial_t+\epsilon \partial_T$, while $d^2/d\tau^2$ (up to the first order in $\epsilon$) takes the form $\partial^2_t+2\epsilon \partial_{t,T}$. With the above rules the zero-th order of the Duffing equation reduces to $\partial^2_t y_0+y_0=0$, which is solved by $y_0(t,T)=R(T)\cos(t+\Phi(T))$. Using $y_0(t,T)$ within the first order equation and imposing vanishing secular terms one obtains the following set of equations for $R(T)$ and $\Phi(T)$:
\begin{eqnarray}
&&2R'+\beta R=f_1\cos(\Phi)+f_2\sin(\Phi)\\\nonumber
&&2\Phi'R+aR-\frac{3}{4}b R^3=-f_1\sin(\Phi)+f_2\cos(\Phi),
\end{eqnarray}
where $f_1=g_1X_1^{\omega}\cos(\varphi)+g_2 X_2^{\omega}$ and $f_2=g_1 X_1^{\omega}\sin(\varphi)$ ( here we fixed $\varphi_1=0$ and
$\varphi_2=\varphi$), while $R'$ and $\Phi'$ represent the first derivatives with respect to $T$. The above equations completely characterize  the zero-th order solution of the problem allowing to derive the pumping current produced when a weak nonlinear dynamics is present. When the pumping terms are absent ($f_1=f_2=0$) the stationary solution is $y_e=0$  and the static scatterer condition is reached. When non-vanishing pumping terms are present the equilibrium condition $R'=0,\ \Phi'=0$ can be written as follows:
\begin{eqnarray}
\label{eq:equilibrium}
R_{e}^2\Bigl\{\beta^2+\Bigl[a-\frac{3}{4}bR_{e}^2\Bigl]^2\Bigl\}=f_1^2+f_2^2\\\nonumber
\left[
  \begin{array}{c}
    \cos(\Phi_e) \\
    \sin(\Phi_e) \\
  \end{array}
\right]=\mathcal{M}\left[
          \begin{array}{c}
            \frac{\beta R_e}{\sqrt{f_1^2+f_2^2}} \\
            \frac{-a R_e+(3/4)b R_e^3}{\sqrt{f_1^2+f_2^2}} \\
          \end{array}
        \right],
\end{eqnarray}
where $\mathcal{M}=(\mathbb{I}f_1-i\sigma_{y}f_2)/\sqrt{f^2_1+f^2_2}$ is a rotation matrix, being $\sigma_y$ a Pauli matrix. Once the equilibrium radius $R_e$ is known solving the first equation in (\ref{eq:equilibrium}) the dynamical phase shift $\Phi_e$ is determined by the second equation in (\ref{eq:equilibrium}). In this way the asymptotic zero-th order solution is completely determined, $y_e(\tau)=R_e\cos(\tau+\Phi_e)$. Using the asymptotic form of the scattering matrix $\mathcal{S}(X_1(\tau),X_2(\tau); y_e(\tau))$ and proceeding as in the standard case,  one obtains the formula of the pumped current:
\begin{eqnarray}
\label{eq:main_result}
I_{\alpha}=I^{0}_{\alpha}&+&\frac{e\omega}{2\pi}\sum_{\beta,j}\{ \cos(\Phi_e)[R_e X_j^{\omega}\Lambda^{j}_{\alpha\beta}\cos(\varphi_j)]+\nonumber \\
&+&\sin(\Phi_e)[R_e X_j^{\omega}\Lambda^{j}_{\alpha\beta}\sin(\varphi_j)]\},
\end{eqnarray}
where $\Lambda^{j}_{\alpha\beta}=\Im\{(\partial_y\mathcal{S}_{\alpha\beta}\partial_{X_j}\mathcal{S}^{\ast}_{\alpha\beta})_0\}$, while $I_{\alpha}^0$ is the standard pumping term as given in Eq.(\ref{eq:pump_curr_0}). Eq.(\ref{eq:main_result}) represents the main result of this work. Its validity goes beyond the Duffing dynamics we are discussing here. Indeed every weakly nonlinear problem of the form
\begin{equation}
\ddot{x}+x+\epsilon h(x,\dot{x},\tau)=0
\end{equation}
has a zero-th order asymptotic solution of the form $x_e(\tau)=R_e \cos(\tau+\Phi_e)$, being $R_e$ and $\Phi_e$ determined by the precise form of the nonlinear time-periodic function $h(x,\dot{x},\tau)$. Thus Eq.(\ref{eq:main_result}) has to be considered valid for the whole class of weakly nonlinear problems. From the experimental point of view, the parasitic nonlinear dynamics of the variable $y_e(\tau)$ can be probed using Eq.(\ref{eq:main_result}) and considering the quantities $R_e\sin(\Phi_e)\Lambda^{j}_{\alpha\beta}$ and $R_e\cos(\Phi_e)\Lambda^{j}_{\alpha\beta}$ as fitting parameters. For the fitting procedure it can be useful to estimate $\partial_y\mathcal{S}_{\alpha\beta}$ by using the Fisher-Lee relation
\begin{equation}
\label{eq:fisher-lee}
\mathcal{S}_{\alpha\beta}(E_F,\tau)=\delta_{\alpha\beta}-i\sqrt{\Gamma_{\alpha}(X_1,X_2)\Gamma_{\beta}(X_1,X_2)}G^r(E_F,\tau),
\end{equation}
 where  $G^r(E_F,\tau)$ is the instantaneous retarded Green's function associated to the scattering region assumed of resonant-like form
 \begin{equation}
 G^r(E_F,\tau)=[E_F-\varepsilon(y_e(\tau))+i\sum_{s}\Gamma_s(X_1(\tau),X_2(\tau))/2]^{-1},
 \end{equation}
 while the tunneling amplitudes $\Gamma_s$  are related to the Green's function self-energy. In this way we obtain a model for the parametric derivative of the scattering matrix with respect to the parasitic variable:
\begin{equation}
(\partial_y\mathcal{S}_{\alpha\beta})_0=\frac{-i\lambda\sqrt{\Gamma^0_{\alpha}\Gamma^0_{\beta}}\exp(-2i\Theta)}{(E_F-\varepsilon_0)^2+(\sum_s\Gamma_s^{0}/2)^2},
\end{equation}
where $\lambda=(\partial_y\varepsilon(y))_0$, $\Theta=\arctan([\sum_s\Gamma_s^{0}/2]/[E_F-\varepsilon_0])$, $\varepsilon_0$ being  the energy level of the resonant state inside the scatterer in the absence of pumping terms. When $\varepsilon_0 \approx E_F$ and considering $\Gamma^{0}_L \approx \Gamma^{0}_R=\Gamma_0$, one obtains the simple relation $(\partial_y\mathcal{S}_{\alpha\beta})_0 \approx i\lambda/\Gamma_0$ leading to the useful estimation $\Lambda^{j}_{\alpha\beta} \approx (\lambda/\Gamma_0)\Re\{(\partial_{X_j}\mathcal{S}_{\alpha\beta})_0\}$. Within the resonant hypothesis considered above the nonlinear dynamics affects the pumped current mainly via the terms $R_e\sin(\Phi_e)$ and $R_e\cos(\Phi_e)$.
Thus in case of bistability of the forced nonlinear dynamics an hysteresis of the pumped current induced by $R_e$ is expected. In the latter case, the quantum pump displays a memory effect controlled by the features of the pumping cycle. To explain this point we now go back to the analysis of the forced Duffing oscillator governing in our specific example the dynamics of $y$.\\

\section{Results for the Duffing Oscillator}
\label{sec:results}
\begin{figure}
\centering
\includegraphics[scale=.35]{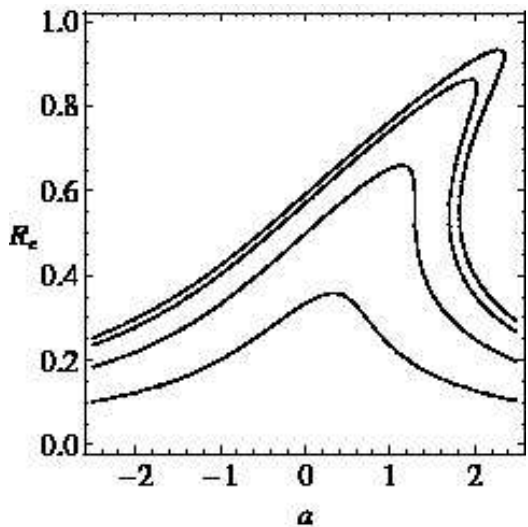}\\
\caption{Equilibrium radius $R_e$ as a function of the detuning parameter $a$ by setting the remaining parameters as follows:
$\beta=0.75$, $b=3.5$, $g_1X_1^{\omega}=g_2X_2^{\omega}=0.35$. The curves have been obtained from top to bottom by setting the pumping phase as $\varphi=0$, $\pi/4$,  $\pi/2$,  $3\pi/4$.} \label{fig:fig1}
\end{figure}

 The numerical analysis of first equation in (\ref{eq:equilibrium}) is shown in Fig.\ref{fig:fig1}, where the equilibrium radius $R_e$ is reported as a function of the detuning parameter $a$ and for different pumping phases $\varphi$. Starting from the upper curve obtained for $\varphi=0$ we observe an hysteretic behavior as a function of $a$. Increasing the value of the pumping phase the hysteresis disappears
when the value $\varphi \approx \pi/2$ is reached.
The hysteretic behavior of the Duffing oscillator is controlled by the nonlinear term $b$. In particular for values of $b$ below the critical value $b_c$ the curves $R_e(a)$ are single valued functions, while when $b_c$ is exceeded (i.e. $b>b_c$) a bistable behavior is detected. Since in our case $b=3.5$ is fixed, we have to assume that the pumping phase $\varphi$ can vary $b_c$. Very interestingly the above observation implies that the shape of the pumping cycle may affect directly the stability of the nonlinear parasitic variable. In particular the hysteresis indicates that the phase space of the oscillator contains two stable limit cycles characterized by different radius $R_e^{(1)}$ and $R_e^{(2)}$, a third unstable limit cycle being  present between the stables one.  In this way the state of the parasitic variable can be controlled by using the detuning $a$ (or equivalently the pumping frequency $\omega$) and the pumping phase $\varphi$. Furthermore, the change of state of the parasitic variable produces an observable variation of the pumping current allowing us to modify and detect the state of the parasitic variable $y$. In this way a quantum-pump-based memory device is obtained.
\begin{figure}[t]
\centering
\includegraphics[scale=.85]{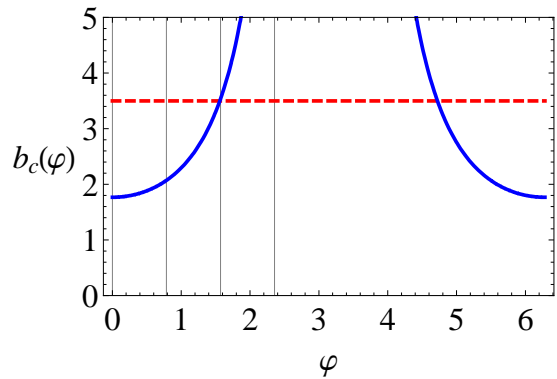}\\
\caption{Critical value (full line) of the nonlinear term $b_c$ as a function of the pumping phase $\varphi$ setting the remaining parameters as follows: $\beta=0.75$, $b=3.5$, $g_1X_1^{\omega}=g_2X_2^{\omega}=0.35$. The dashed line represents the the actual value of the nonlinear term $b=3.5$. When the full line lies below the dashed one the condition $b>b_c$ is fulfilled and an hysteretic behavior is detected in Fig.\ref{fig:fig1}. For $b<b_c$, $R_e(a)$ is a single valued function. The vertical lines correspond to the pumping phases chosen in Fig.\ref{fig:fig1}.} \label{fig:fig2}
\end{figure}
The critical value $b_c$ of the nonlinear term can be computed exactly differentiating the first relation in (\ref{eq:equilibrium}) with respect to $a$ and imposing diverging derivative for $R_e(a)$ (i.e. $d R_e/d a \rightarrow \pm \infty $). After some additional algebra, one obtains the following expression for $b_c(\varphi)$:
\begin{equation}
b_c(\varphi)=\frac{32\sqrt{3}\beta^3}{27[f_1^2+f_2^2]},
\end{equation}
where the dependence on the pumping phase $\varphi$ is hidden in the forcing terms $f_1$ and $f_2$. The critical curve $b_c(\varphi)$ is reported in Fig.\ref{fig:fig2} as a function of the pumping phase setting the remaining parameters as done in Fig.\ref{fig:fig1}. The dashed line represents the actual value of the nonlinear term (i.e. $b=3.5$).
The analysis of the figure shows that when the critical curve $b_c(\varphi)$ lies below the dashed line (i.e. when $b>b_c$) an hysteretic behavior affects the system, while an increasing of the pumping phase  beyond $\varphi \approx \pi/2$ destroys the bistability.
Thus we can manipulate the hysteretic threshold  $b_c$ by modifying the features of the pumping cycle (i.e. $X_i^{\omega}$, $\varphi$).\\

\subsection{A quantum-pump-based memory device}
\begin{figure}
\centering
\includegraphics[scale=.5]{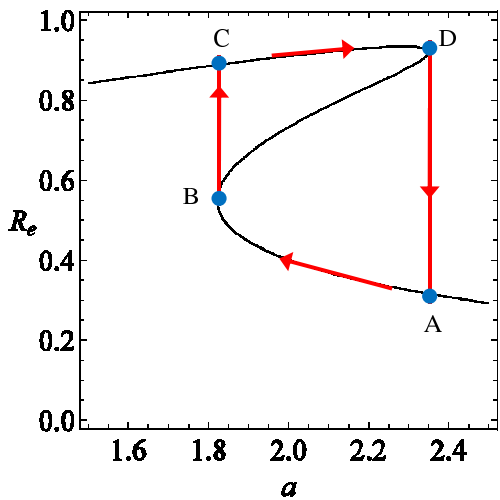}\\
\caption{Hysteresis loop for the radius $R_e$ as a function of the detuning parameter $a$ by setting the remaining parameters as follows:
$\beta=0.75$, $b=3.5$, $g_1X_1^{\omega}=g_2X_2^{\omega}=0.35$, $\varphi=0$.} \label{fig:fig3}
\end{figure}
 In this Section we discuss how a memory device based on the quantum pumping works. In order to build a memory device a scattering region containing a nonlinear parasitic dynamics is needed. For instance, a parasitic dynamics can be activated  by an unknown capacitive coupling between the scatterer and a charge density located on the  mesoscopic substrate which can be usually modeled by a classical RLC circuit. Another more controllable way to produce nonlinear parasitic dynamics is provided by a scatterer able to elastically react in the presence of a charge density on it. Such system is a nanoresonator coupled to external leads whose experimental realization is reported in Ref.[\onlinecite{sazonova04}], while its use as deformable quantum dot in the quantum pumping context has been reported in Ref.[\onlinecite{deformable-qd}]. Following a derivation similar to the one given in Ref.[\onlinecite{deformable-qd}] and going beyond the linear terms in expanding the charge density, one obtains the center of mass dynamics of the scatterer (i.e. the parasitic degree of freedom dynamics) in the form given in Eq.(\ref{eq:duffing_eom}). Thus Eq.(\ref{eq:duffing_eom}) has to be considered the minimal nonlinear model of the center of mass dynamics, while the term $\epsilon[-a y+b y^3+\sum_{i=1,2}g_iX_i^{\omega}\sin(\tau +\varphi_i)]$ has to be meant as the electrostatic force acting on the nanoresonator with the capacitive coupling provided by $\epsilon$. Once the center of mass dynamics has been recognized in the form of the Duffing oscillator all the arguments concerning its stability are the same as discussed in Sec.\ref{sec:results}. In particular, we now study the case in which the system presents a bistable behavior which is needed to obtain a memory device.  In Fig.\ref{fig:fig3} we present the hysteresis loop obtained setting the system parameters as follows: $\beta=0.75$, $b=3.5$, $g_1X_1^{\omega}=g_2X_2^{\omega}=0.35$, $\varphi=0$. It represents the equilibrium radius $R_e$ as a function of the detuning parameter $a$ which can be experimentally controlled  by using gate voltages or acting on the frequency of the pumping cycle. Starting from the point $A$ of the loop and decreasing the value of $a$, we move along the lower branch until the point $B$ is reached. As evident, from analyzing the second and third term in Eq.(\ref{eq:main_result}), along $AB$ the system presents a continuous variation of the pumping current since $R_e$ and $\Phi_e$ are continuous function of $a$. Further decreasing $a$, moving from the point $B$, produces a jump toward the point $C$ located on the upper branch (i.e. $CD$) of the hysteresis loop. This jump, corresponding to a sizeable change of $R_e$, can be detected by a strong variation in the pumping current if the second and third terms in Eq.(\ref{eq:main_result}) are of the same order of magnitude compared to the pumping term. Once the point $C$ is reached, the loop can be closed by increasing $a$ until the state $D$ is achieved. Staring from $D$, a further increasing of the detuning $a$ produces a big jump toward the initial state $A$ of the hysteresis loop. The $DA$ jump (as also the $BC$ jump), detectable in the pumping current, indicates the change of state of the parasitic dynamics (e.g. the center of mass evolution of the nanoresonator). In this way the bistable state of the parasitic variable is recorded by two reference pumping currents taking different values pertaining to the upper or lower branch of the hysteresis loop. The above procedure constitutes the working principle of a memory device based on quantum pumping.

\section{Conclusions}
\label{sec:conclusions}
 In conclusion, we have shown an unconventional application of the quantum pumping in which a parasitic nonlinear variable affects the pumping cycle introducing interesting memory effects controlled by the critical parameter $b_c$. The hysteretic threshold $b_c$ can be manipulated acting on the shape of the pumping cycle and the state of the parasitic variable $y$ can be detected measuring the pumping current. The features above are useful in obtaining a memory device based on quantum pumping which could be realized using a nano-resonator similar to the one reported in Ref.[\onlinecite{sazonova04}].
Apart from the technological application of our study, nanoelectromechanical
systems are ideal candidates to validate the present theory
and explore the parasitic dynamics.
Studies of nanoelectromechanical systems are usually performed by reducing
their dynamics from that of a continuous model to that of the center of mass
dynamics, the so called "resonator". However in the real situation the resonator
is an extended object whose center of mass motion is affected by different vibrational modes of the structure. Thus, due to the complexity of
the problem is not possible to fix a priori the parameters of the reduced model (e.g. the damping coefficient, the
resonance frequency, etc) or recognize the structure of
the nonlinear contributions to the damping term. In fact the total energy is differently stored among the vibrational degrees of freedom depending on the working point of the resonator.
Thus, even though the resonator dynamics can be taken
into account in designing the device, to fix
the time evolution of the mechanical degrees of freedom
a precise knowledge of the capacitive coupling as a function of the
resonator position is needed.
Our proposed method can be useful in identifying the model for the capacitive coupling and offer a method to probe the parasitic nonlinear dynamics.\\
 A further example of parasitic dynamics is constituted by the capacitive coupling of a classical background charge to the scattering region. Such situation can be easily understood as the capacitive coupling between the scattering region and an effective RLC circuit. However, also in this simple case, the parasitic dynamics can be imagined as the linearization of some nonlinear model around the working point whose dynamical parameters (capacitance, differential resistance, etc) can be tested by using our theory.\\
Finally, a variety of other useful applications
such as frequency synchronization, frequency mixing and
conversion, and parametric amplification, could be implemented
by our proposal playing an important role for nanomechanical
devices.

\bibliographystyle{prsty}

\begin{thebibliography}{99}
\bibitem{nota1}More realistic models should also include nonlinear damping terms like $(\dot{y})^3$ or $y^2\dot{y}$ omitted here to simplify the theory.
\bibitem{nota2} Notice that the choice $b>0$ does not allow a chaotic behavior of the system.
\end{thebibliography}


\end{document}